**Highlights**

- Proposed a CSRNet-based method for automated honey bee density estimation.
- The method was developed and validated using real field experiments in active hives.
- Processed each image in seconds, enabling fast, large-scale population analysis.
- Applies AI-based methods to improve accuracy and efficiency in honey bee population research.

Fast, accurate measurement of the worker populations of honey bee colonies using deep learning

Junmin Zhong[1], Jon F. Harrison[2], Jennie Si[1], and Jun Chen[2,3*]

[1]School of Electrical, Computer and Energy Engineering, Arizona State University, Tempe, AZ 85281, USA.

[2]School of Life Sciences, Arizona State University, Tempe, AZ 85281, USA.

[3]Department of Mathematics, Texas A&M University - Kingsville, Kingsville, TX 78363, USA.

*Corresponding author. june.chan6193@gmail.com

**Abstract**

Honey bees play a crucial role in pollination, contributing significantly to global agriculture and ecosystems. Accurately estimating hive populations is essential for understanding the effects of environmental factors on bee colonies, yet traditional methods of counting bees are time-consuming, labor-intensive, and prone to human error, particularly in large-scale studies. In this paper, we present a deep learning-based solution for automating bee population counting using CSRNet and introduce ASUBEE, the FIRST high-resolution dataset specifically designed for this task. Our method employs density map estimation to predict bee populations, effectively addressing challenges such as occlusion and overlapping bees that are common in hive monitoring. We demonstrate that CSRNet achieves superior performance in terms of time efficiency, with a computation time of just 1 second per image, while delivering accurate counts even in complex and densely populated hive scenarios. Our findings show that deep learning approaches like CSRNet can dramatically enhance the efficiency of hive population assessments, providing a valuable tool for researchers and beekeepers alike. This work marks a significant advancement in applying AI technologies to ecological research, offering scalable and precise monitoring solutions for honey bee populations.

## 1. Introduction

The number of workers in a honey bee colony is an important index of colony health, and can be predictive of the capacity of the hive to accumulate honey, pollinate, reproduce, and survive stressors (Chabert et al., 2021; Döke et al., 2019; Requier et al., 2017; Seeley, 2014; Delaplane et al., 2013; Donaldson-Matasci et al., 2013; Winston, 1991; Farrar, 1937). Additionally, the number of workers is an important parameter for models attempting to accurately predict environmental, disease, and management effects on honey bee colonies (Chen et al., 2023; Messan et al., 2021; Kuan et al., 2018; Becher et al., 2014; Becher et al., 2013; DeGrandi-Hoffman and Curry, 2004; Bromenshenk and DeGrandi-Hoffman, 1993; DeGrandi-Hoffman et al., 1989). The process of quantifying bee colonies is inherently estimative, and the method used varies depending on the goal. For rapid assessments with minimal disruption to the colony, (Chabert et al., 2021) verified a method of counting the number of interframe spaces occupied by bees without removing the frames. However, their predictive equations had $r^2$ values ranging from 0.54 to 0.81 depending on the specific method, so this method is likely better for seasonal or cross-apiary comparisons rather than for precise quantitative measurements of changes in colony demography over time. A likely more accurate but somewhat subjective method involves pulling frames from hives and having observers estimate the percent of each frame covered by bees (Liebefeld method (Dainat et al., 2020)). Farrar (1937) developed a method that remains in common use (e.g., Capela et al., (2023); each frame is removed from the hive at night, weighed, bees are brushed off, and the frame is reweighed. Then the number of bees is estimated from the average mass of a subset of bees. This has been the most accurate method available, though because of the disruption to the colony, it is not recommended for regular use, limiting the utility of this method for finer-scale assessment of demography (Delaplane et al., 2013). A method that requires less disruption is to pull each frame, photograph it, trace the area occupied by bees, and convert this to bee number based on separate estimates of average bee density (Cornelissen et al., 2009). However, because bee density can be quite variable, this method has some inherent inaccuracies.

Methods for assessing entire bee populations remain time-consuming and often impractical for large-scale studies. For instance, Fisher II et al. (2021) used a camera to photograph each frame and estimated the number of adult worker bees using ImageJ (National Institutes of Health, Bethesda, MD). In a typical colony of 7-10 frames, with 2 sides per frame, the counting process becomes labor-intensive, especially when scaling to 40 colonies in an experiment. Similarly, Minucci et al. (2021) estimated population and food storage by converting area measurements into individual or cell counts, while DeGrandi-Hoffman et al. (2023) counted bees by overlaying a grid on the frame and summing the squares with bee coverage. With the rise of computer vision technologies, there is significant potential for deep learning-based techniques to automate and improve the accuracy of bee counting from digital photographs. Crowd counting is a computer vision task that aims to estimate the number of individuals in crowded scenes from images or videos. Crowd counting methods are generally categorized into detection-based, regression-based, and point-supervision-based approaches. Detection-based methods, like those in (Liu et al., 2019; Liu et al., 2018), predict bounding boxes to count individuals, but struggle with occlusion in dense areas and require extensive annotations. To address these limitations, regression-based approaches (Chan et al., 2008) were developed, initially focusing on global image counts but lacking spatial context. The introduction of density maps (Lempitsky and Zisserman, 2010), along with techniques like multi-scale mechanisms (Ma et al., 2020; Cao et al., 2018), perspective estimation (Yang et al., 2020; Yan et al., 2019), and auxiliary tasks (Liu et al., 2020; Zhao et al., 2019), significantly advanced these methods. More recently, point-supervision approaches, such as BL (Ma, Wei, Hong, and Gong 2019), have improved accuracy by avoiding pseudo-map generation, with other works exploring optimal transport and divergence measures (Ma, Wei, Hong, Lin, et al., 2021; Wang et al., 2020).

There are several key obstacles to doing the task: 1) There is no existing image dataset specifically designed for training crowd counting methods in this context. 2) No crowd counting method is currently tailored to our task. Given the significant variations in the scale of bees across images, we must incorporate multi-scale features to accurately estimate crowd counts in diverse images. In this paper, we aim to achieve accurate crowd counting from arbitrary still images under different daytime conditions and varying crowd densities (see Figure 1 for examples). The contributions of this paper are as follows:

1) We introduce a new dataset, ASUBEE, specifically designed for training crowd counting models for bee population assessment. ASUBEE consists of nearly 400 high-resolution images (5184*3456 resolution) with approximately 350,000 accurately labeled bees.
2) We develop a complete training and deployment framework tailored to our task, based on the well-established crowd counting algorithm, CSRNet (Li et al., 2018).

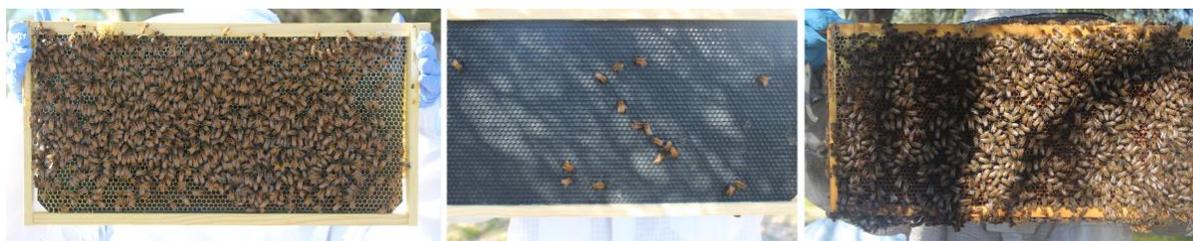

Fig. 1: Representative images of our ASUBEE dataset. Left: a dense cluster of bees in soft natural light. Middle: a small group of bees in shaded areas. Right: a crowded gathering of bees, with shading and overlapping.

2. Materials and Methods

Since existing datasets are unsuitable for the crowd counting task addressed in this work, we introduce a new large-scale bee crowd counting dataset, ASUBEE. This dataset comprises 370 annotated images, totaling 353,388 bees, with the center of each bee's body annotated. To the best of our knowledge, ASUBEE is the only dataset specifically designed for crowd bee counting.

2.1. Image collection and annotation

The hive was carefully opened, and each frame was gently handled to ensure the adult worker bees remained on the frame. A Canon® EOS Rebel T5 camera was used to capture photographs of both sides of each frame, and this process was repeated until all frames in the hive were photographed. Each hive contained either one or two boxes, with up to 10 frames per box. For three days within a week, photographs were taken for 40 hives. Image J software (NIH, Research Triangle Park, NC) was used to quantify the adult worker bee population in each photograph through point-counting. The position of each point, representing an individual bee, was stored as a label. Each point was randomly placed at different locations on the bee's body.

2.2. Training and testing set

Due to the high resolution of our raw images (5184x3456 pixels), which is too large for training deep neural networks, we cropped each original image into 24 sub-images of 224x224 pixels. This process resulted in a total of 8,880 sub-images, which we divided into training and testing sets: 7,104 images for training and 1,776 for testing. Table 1 provides the statistics of the ASUBEE dataset. If this work is accepted for publication, we will release the dataset, annotations, and the full training and deployment codes.

|  | **Training** | **Testing** |
|---|---|---|
| Number of images | 7104 | 1776 |
| Maximal number of **bees/image** | 212 | 154 |
| Minimal number of **bees/image** | 0 | 0 |
| Mean number of **bees/image** | 30.3 | 30.2 |
| Total | 283,017 | 70,370 |

Table 1: Labeled Bees Dataset Statistics.

2.3. Density map

To estimate the number of individuals in an image using convolutional neural networks (CNNs), two common approaches are typically considered. The first approach involves a network that directly outputs the estimated count from the image input. The second approach generates a density map, showing the distribution of individuals across the image (e.g., how many bees per square centimeter), with the total count derived by integrating the density map (Li et al., 2018; Zhang et al., 2016). In this paper, we chose to use density mapping as this approach preserves more information. Compared to a single overall count, a density map provides the spatial distribution of the crowd within the image, which is useful in many contexts. For example, if only part of a bee is captured in the image, the density map assigns a probability of less than one, preventing over-counting. Additionally, as shown in Fig. 1 (right), the lower-left and lower-right corners display clusters of bees, and by using a density map, the CNN can detect potential irregularities in these areas, improving the accuracy of the count. Also, density mapping allows better adaptation of the model to varying scales. When learning a density map through a CNN, the filters naturally adjust to detect heads (or objects) of various sizes, making them more effective for diverse inputs with significant changes in perspective. This adaptability enhances the semantic meaning of the filters, leading to improved accuracy in crowd counting. This capability is particularly important in our case, as images are collected from varying distances and angles, which introduces a wide range of scale and perspective variations.

To train a CNN to estimate crowd density from an input image, the quality of the density maps used for training is crucial. The process begins by converting labeled heads in an image into a crowd density map. A head at pixel location $x_i$ is represented as a delta function $\delta(x - x_i)$, and the image with $J$ heads is represented as the sum of these functions:

$$H(x) = \sum_{i=1}^{J} \delta(x - x_i), \tag{1}$$

To transform this into a continuous density function, it is convolved with a Gaussian kernel (Lempitsky and Zisserman 2010) $G_\sigma$, $\sigma$ indicating standard deviation in a Gaussian kernel:

$$F(x) = H(x) \times G_\sigma(x), \tag{2}$$

In our study, a kernel size of 15 and $\sigma = 4$ gives the best result. In Fig. 2, we have shown the so-obtained density maps of two exemplar images in our dataset.

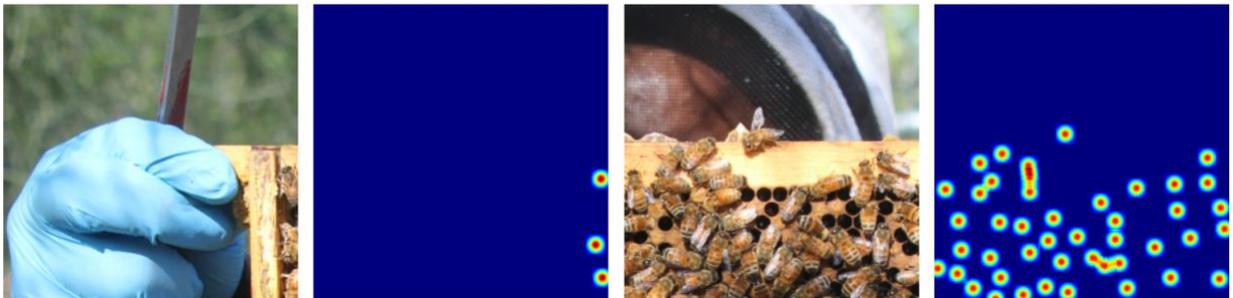

Fig. 2. Original images and corresponding crowd density maps obtained by the Gaussian kernels.

### 2.4. Counting Method: CSRNet Architecture

In this work, we use a well-established crowd counting method, CSRNet (Li et al., 2018). The fundamental idea of CSRNet is to deploy a deeper CNN to capture high-level features with larger receptive fields, thereby generating high-quality density maps without excessively increasing the network complexity. In this section, we first introduce the architecture of CSRNet, followed by the corresponding training methods.

As Figure 3 shows, CSRNet leverages the VGG-16 (Simonyan and Zisserman, 2014) architecture as its front-end to capture high-level features with strong transfer learning capabilities. By removing the fully connected layers from VGG-16, CSRNet utilizes only the convolutional layers, allowing for effective feature extraction without overly shrinking the resolution of the output. The front-end reduces the output size to 1 of the original image, preserving spatial information crucial for density estimation. To avoid further reduction in resolution, which would hinder the generation of accurate density maps, CSRNet incorporates dilated convolutional layers in the back-end. These dilated layers allow the network to extract deeper contextual information while maintaining the output resolution, leading to high-quality density maps. This architecture is designed to balance depth and computational efficiency, enabling accurate crowd counting without the complexity of excessive network layers.

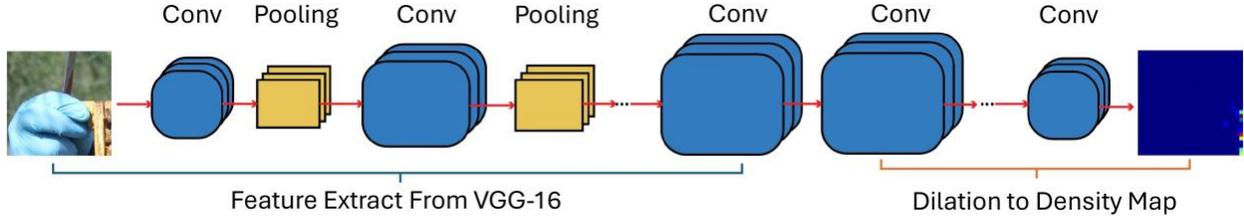

Fig. 3. Structure of CSRNet. Conv is the convolution layer, and Pooling is the max-pooling layer. Convolution detects patterns like edges, while max-pooling downsamples by keeping the strongest signals. Convolution and pooling layers act as feature extractors from the input image, while dilated convolution expands the area each convolution filter can cover, enabling it to extract more global or context-aware features from an image while maintaining the spatial resolution of the feature map.

The detailed network configuration of CSRNet is shown in Table A.1 in Appendix A. The first 10 convolutional layers are taken from the pre-trained VGG-16 model (Simonyan and Zisserman, 2014). For the remaining layers, the weights are initialized using a Gaussian distribution with a standard deviation of 0.01. The model is trained using the stochastic gradient descent (SGD) optimizer with a fixed learning rate of $1 \times 10^{-7}$. The loss function used is the Euclidean distance, which measures the difference between the ground truth and the predicted density map. The loss function $L(\theta)$ is defined as:

$$L(\theta) = \frac{1}{2N}\sum_{i=1}^{N}\|Z(X_i; \theta) - Z_i^*\|_2^2, \qquad (3)$$

where $N$ is the size of the training batch, and $Z(X_i; \theta)$ represents the output generated by CSRNet with parameters $\theta$. $X_i$ denotes the input image, and $Z_i^*$ is the ground truth density map corresponding to the input image $X_i$.

2.5. Hyperparameters

The training hyperparameters are outlined in Table 2. An epoch refers to a full pass through the entire training dataset by the learning algorithm. During each epoch, the model processes every example in the dataset once, adjusting its parameters (weights and biases) based on the error calculated using Eq. (3). The batch size used in Eq. (3) consists of 4 samples. We apply the SGD optimizer to update the model parameters, with the key hyperparameters for SGD configured as shown in Table 2.

| Hyperparameter | Value |
|---|---|
| Max Epoch | 200 |
| Batch size $N$ | 4 |
| SGD momentum | 0.95 |
| SGD Weight decay | $1 \times 10^{-4}$ |
| SGD Learning rate | $1 \times 10^{-7}$ |

Table 2: Hyperparameters used by the SAC algorithm

2.6. Evaluation metrics

The Mean Absolute Error (MAE) and Mean Squared Error (MSE) are used for evaluation, defined as:

$$MAE = \frac{1}{N}\sum_{i=1}^{N}|C_i - C_i^*|, \qquad (4)$$

$$MSE = \sqrt{\frac{1}{N}\sum_{i=1}^{N}(C_i - C_i^*)^2}, \qquad (5)$$

where $N$ is the number of images in the test sequence, and $C_i^*$ is the ground truth count. $C_i$ represents the estimated count, defined as follows:

$$C_i = \sum_{l=1}^{L}\sum_{w=1}^{W} z_{l,w}. \qquad (6)$$

Here, $L$ and $W$ denote the length and width of the density map, respectively, and $z_{l,w}$ is the pixel value at position $(l, w)$ of the generated density map. Therefore, $C_i$ is the estimated counting number for image $X_i$.

We also use the error rate to evaluate the quality of the counting. The error rate $\epsilon$ is defined as:

$$\epsilon = \frac{\sum_{i=1}^{N}|C_i - C_i^*|}{\sum_{i=1}^{N}C_i^*} \qquad (7)$$

and the lower the error rate, the better.

3. Results

The accuracy of the CSRNet model improves along the training process (Fig. 4). The loss curve (see Eq. (3)) exhibits a sharp decline during the initial epochs, and as training progresses, the loss continues to decrease, albeit at a slower rate, suggesting gradual performance improvements as the model nears convergence. Similarly, the MAE curve (see Eq. (4)) follows a steep decline in the early epochs, and as training advances, the MAE stabilizes with only minor fluctuations, signifying that the model is refining its predictions. Overall, both curves reflect successful learning, with consistent decreases in both loss and MAE, demonstrating that the model effectively minimizes errors and achieves increasing accuracy as the number of epochs grows.

Examples of model output for three test images are shown in Fig. 5. In the first sample, where the human count is 11, the model predicts a count of 11.31 with an error rate of 0.028, showing that the model's prediction closely matches the ground truth. In more densely populated bee samples, such as the second example with a human count of 115, CSRNet estimates the count to be 112.6, achieving an even lower error rate of 0.021. Remarkably, even in challenging cases where heavy shading makes it difficult for the human eye to distinguish bees—such as the third sample, which has a count of 81—the model still provides a highly accurate estimate of 84.36, maintaining a low error rate of 0.041. This demonstrates the model's robustness in handling complex visual conditions.

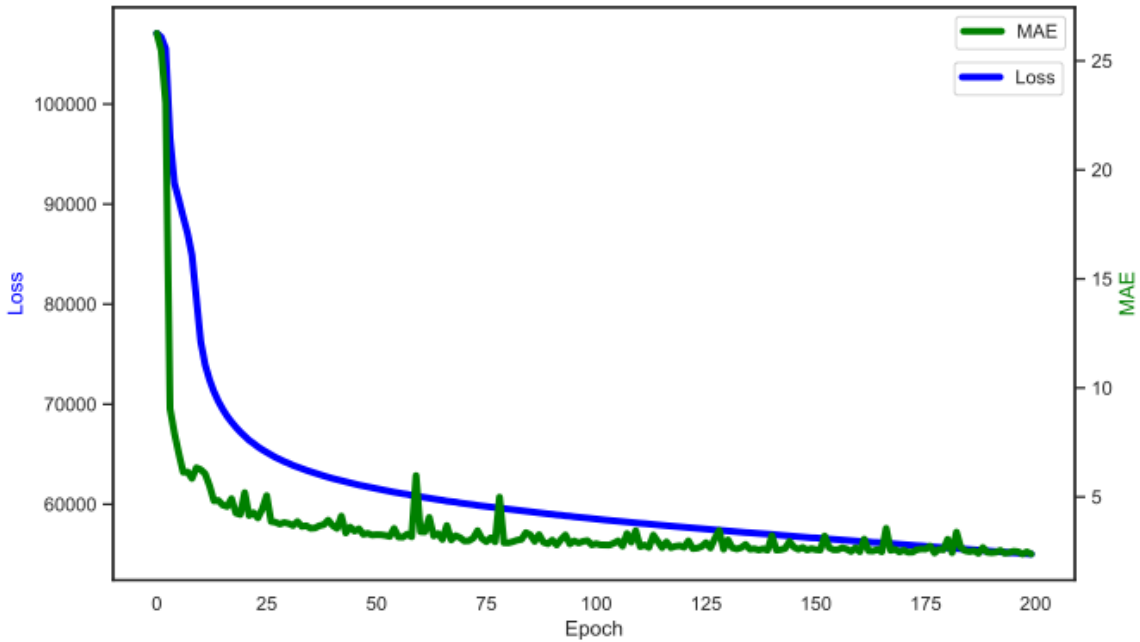

Fig. 4. Training plots of CSRNet on ASUBEE dataset. The x-axis is the training epoch, and y-axis is the training loss (blue) and MAE (green).

3.1. Calculating Whole-Colony Worker Population

To calculate the whole-hive worker population, we summed the counts from each image from that hive. To test the ability and accuracy of the crowd counting approach, we tested the trained model on nine new hives (that were not in the training or testing set). The CSRNet model was able to estimate the worker population of hives with high accuracy, demonstrating low error rates across different hives (Table 3). The error rates range from 0.006 to 0.058, indicating that the model can effectively generalize to new hives that were not part of the training or testing sets. This highlights the robustness and reliability of the crowd counting approach in assessing the population of entire hives.

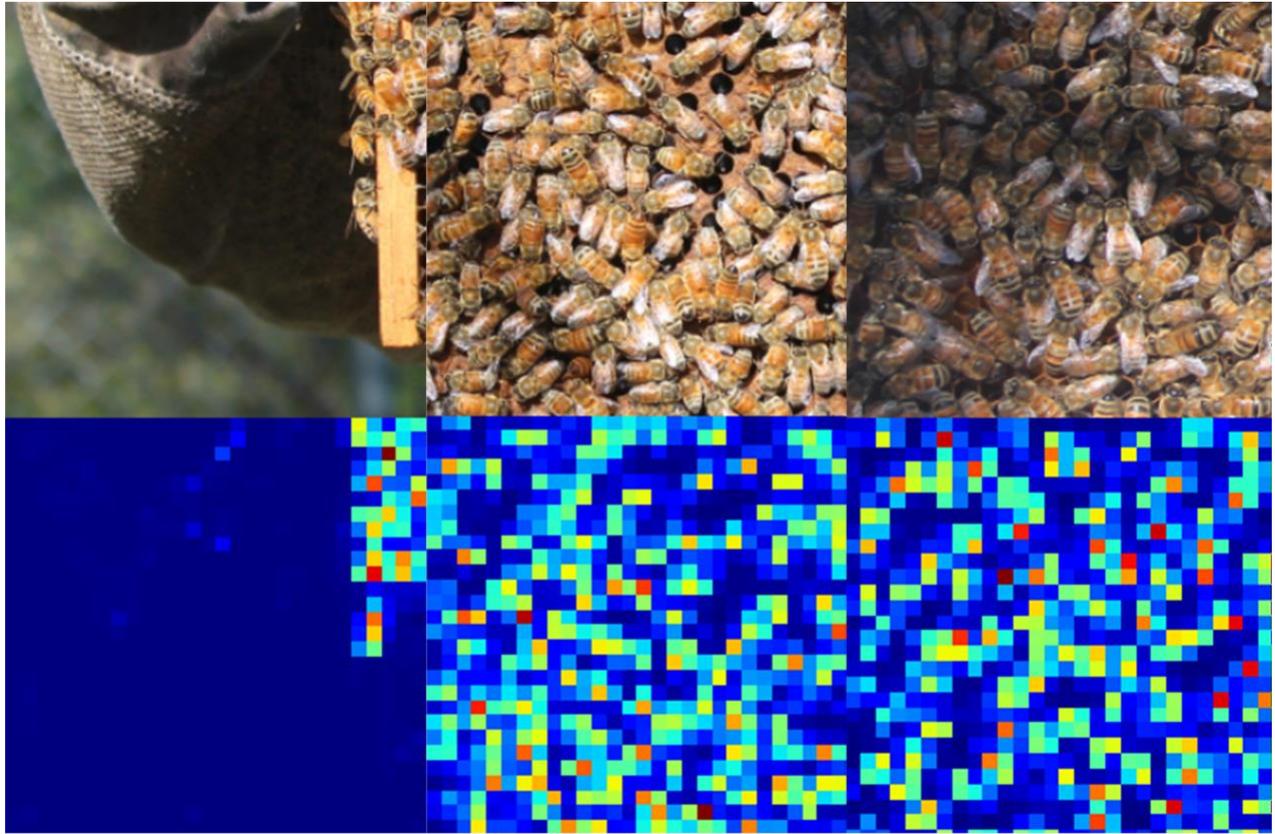

Human Count: 11  
Estimate: 11.31  
Error rate: 0.028  

Human Count: 115  
Estimate: 112.6  
Error rate: 0.021  

Human Count: 81  
Estimate: 84.36  
Error rate: 0.041  

Fig. 5. The first row shows the samples of the testing set in the dataset. The second row presents the generated density map by CSRNet.

| Hive | Manual Count | CSRNet Count | Error Rate |
|------|--------------|--------------|------------|
| 1 | 12485 | 13118 | 0.05 |
| 2 | 11408 | 11798 | 0.03 |
| 3 | 13120 | 13572 | 0.03 |
| 4 | 9258 | 9439 | 0.02 |
| 5 | 12634 | 12714 | 0.006 |
| 6 | 17381 | 16374 | 0.058 |
| 7 | 9739 | 10293 | 0.057 |
| 8 | 10011 | 9429 | 0.058 |
| 9 | 14006 | 13928 | 0.006 |

Table 3: Estimation Errors Based on the whole Hive population

To test the capacity of CSRNet to accurately count workers in images with overlapping bees, we compared CSRNet outputs to estimates made by students and experts of the worker count of three images with clearly overlapping bees (Fig. 6). CSRNet estimates the population by generating a density map. In areas with significant overlap, CSRNet can recognize the crowd density and estimate the total number of bees based on learned patterns, leading to more robust and consistent results. CSRNet consistently produced estimates that fell near the student and expert counts (Table 4).

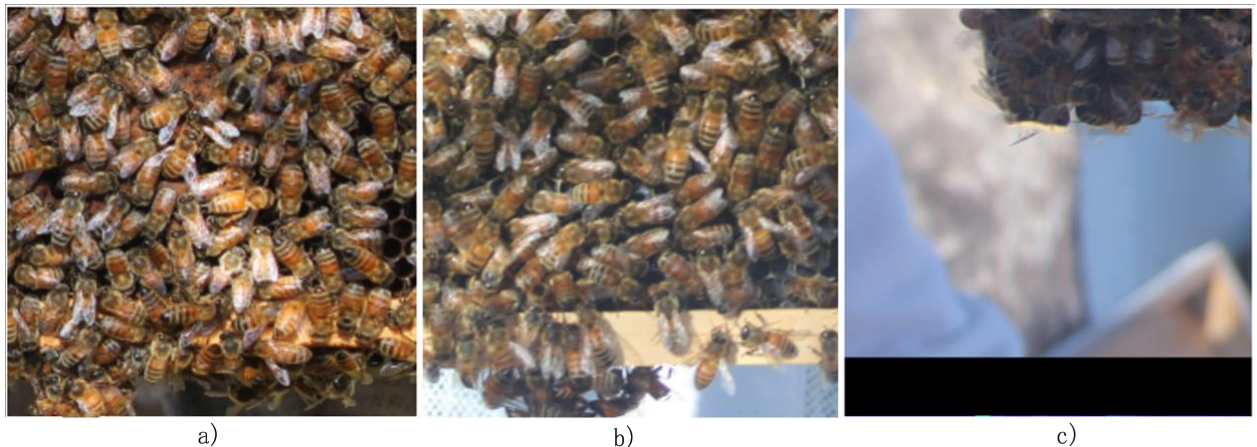

a)          b)          c)

Fig. 6. Examples of figures with overlapping bees.

|  | a | b | c |
|---|---|---|---|
| Student Count | 126 | 90 | 14 |
| Expert Count | 122 | 106 | 20 |
| CSRNet Count | 123.2 | 94.9 | 19 |
| Human Time | 110s | 88s | 60s |
| CSRNet Time | 1.07s | 1.07s | 1.06s |

Table 4: Comparison of human and CSRNet counts on images with overlapping bees. Letters a, b, and c indicate Fig. 6(a), Fig. 6(b), and Fig. 6 (c). "Student count" refers to counts performed by trained students, while "Expert count" represents those done by experienced experts. "CSRNet count" is the output generated by the CSRNet model. "Human Time" refers to the average time taken by humans to complete the count, and "CSRNet Time" is the computation time required by the model to process the image.

4. Discussion

Through our experiments, CSRNet consistently delivered highly accurate counts (generally with errors less than 5%), even in challenging cases with overlapping bees, outperforming human counters in terms of processing time by approximately 100-fold. The low error rates achieved by CSRNet, along with its ability to process images in just over a second, make it a practical and scalable solution for real-time population monitoring in large-scale research projects. Additionally, CSRNet's probabilistic density mapping helps mitigate the impact of occlusion and overlapping bees, providing more reliable results than traditional methods. Our system may not entirely exclude drones from the counts, but the low error rates observed suggest that the model effectively differentiates between worker bees and drones. CSRNet's multi-scale feature incorporation enhances its ability to handle variations in bee size and crowd density, similar to how humans are trained to identify drones based on their distinct shape and size. In the ASUBEE dataset, trained students and experts successfully excluded most drones during annotation. While CSRNet may occasionally include drones, they are typically assigned a low probability in the density map. In healthy hives, where drones generally comprise only 10% to 15% of the population (Winston, 1991), CSRNet achieves very low error rates across nine hives, as shown in Table 3. Additionally, Table 4 demonstrates CSRNet's efficiency, processing an image in approximately 1.07 seconds compared to the 60 to 110 seconds required by humans. While it may not completely eliminate drone counting, CSRNet's low error rates and rapid processing speed make it a highly effective solution for large-scale bee population monitoring.

One concern for this approach is that we took photographs during the day, when a significant number of bees were likely foraging. For a more accurate measure of hive demography, this technique should be done at night. Another negative is that because this method requires the removal of all hive frames, considerable time is required in the apiary, with significant disruption to the hive. For research experiments in which colony disruption must be minimized, X-ray (Greco, 2010) or thermographic measurements (López-Fernández et al., 2018) may eventually provide long-term, non-disruptive monitoring approaches. When relative numbers are all that is required, and errors of 20-40% are tolerable, the method of counting interframe regions with bees without moving the frames is recommended (Chabert et al., 2021).

In summary, our approach represents a significant advancement in automating the bee counting process, offering researchers and beekeepers a robust tool to monitor hive populations more efficiently and accurately. The success of CSRNet in this domain highlights the potential of deep learning technologies to revolutionize ecological research, enabling better understanding and management of bee populations in response to environmental changes. Future work will focus on further refining the model to improve drone exclusion, expanding its application to measurements of other aspects of the hive (honey and pollen stores; eggs, larvae, and pupae), and other pollinator species.

## Author contributions

**Junmin Zhong:** Conceptualization, Methodology, Software, Formal analysis, Data Curation, Visualization, Writing - Original Draft, Writing - review & editing. **Jon F. Harrison**: Conceptualization, Resources, Funding acquisition, Supervision, Writing - review & editing. **Jennie Si**: Resources, Writing - review & editing, Funding acquisition. **Jun Chen:** Conceptualization, Methodology, Validation, Investigation, Data curation, Writing - original draft, Writing - review & editing, Supervision, Project administration.

## Acknowledgments

We would like to thank Alana Boots, Bridget Lachner, Carolyn Marucho, Kylie Maxwell, Pamela Gomez, and Paola Kentel-Rodriguez for helping label the bees. This study was funded by the United States Department of Agriculture – NIFA [2022- 67013-36285] and the National Science Foundation [2211740].

**Appendix A**

| Configurations of CSRNet Input |
| --- |
| VGG-16 FrountEnd: |
| Conv3-64-1 |
| Conv3-64-1 |
| max-pooling |
| Conv3-128-1 |
| Conv3-128-1 |
| max-pooling |
| Conv3-256-1 |
| Conv3-256-1 |
| Conv3-256-1 |
| max-pooling |
| Conv3-512-1 |
| Conv3-512-1 |
| Conv3-512-1 |
| Dilation BackEnd |
| Conv3-512-2 |
| Conv3-512-2 |
| Conv3-512-2 |
| Conv3-256-2 |
| Conv3-128-2 |
| Conv3-64-2 |
| Conv1-1-1 |

Table A.1. Configuration of CSRNet. All convolutional layers use padding to maintain the previous size. The convolutional layers' parameters are denoted as "conv-(kernelsize)-(numberoffilters)(dilationrate)", max-pooling layers are conducted over a 2×2 pixel window with stride 2.